\documentclass[superscriptaddress,nobibnotes,prb, preprint]{revtex4-1}

\usepackage{amsmath,amssymb} 
\usepackage[normalem]{ulem}
\usepackage{braket}
\usepackage{dsfont}
\usepackage{graphicx}
\usepackage{dcolumn}
\usepackage{bm}
\usepackage{color}
\usepackage{upgreek}
\usepackage{multirow}

\usepackage[dvipsnames]{xcolor}

\begin{document}

\title{Two-stage growth for highly ordered epitaxial C$_{60}$ films on Au(111)}

\author{Alexandra B. Tully}\thanks{Equal contributions}
\affiliation{Department of Physics and Astronomy, University of British Columbia, Vancouver, British Columbia, V6T 1Z1 Canada}
\affiliation{Quantum Matter Institute, University of British Columbia, Vancouver, British Columbia, V6T 1Z4 Canada}

\author{Rysa Greenwood}\thanks{Equal contributions}
\affiliation{Department of Physics and Astronomy, University of British Columbia, Vancouver, British Columbia, V6T 1Z1 Canada}
\affiliation{Quantum Matter Institute, University of British Columbia, Vancouver, British Columbia, V6T 1Z4 Canada}

\author{MengXing Na}
\affiliation{Department of Physics and Astronomy, University of British Columbia, Vancouver, British Columbia, V6T 1Z1 Canada}
\affiliation{Quantum Matter Institute, University of British Columbia, Vancouver, British Columbia, V6T 1Z4 Canada}

\author{Vanessa King}
\affiliation{Department of Chemistry, University of British Columbia, Vancouver, British Columbia, V6T 1Z1 Canada}
\affiliation{Quantum Matter Institute, University of British Columbia, Vancouver, British Columbia, V6T 1Z4 Canada}

\author{Erik M\aa rsell}
\affiliation{MAX IV Laboratory, Lund University, Sweden}

\author{Yuran Niu}
\affiliation{MAX IV Laboratory, Lund University, Sweden}

\author{Evangelos Golias}
\affiliation{MAX IV Laboratory, Lund University, Sweden}

\author{Arthur K. Mills}
\affiliation{Quantum Matter Institute, University of British Columbia, Vancouver, British Columbia, V6T 1Z4 Canada}

\author{Giorgio Levy de Castro}
\affiliation{Quantum Matter Institute, University of British Columbia, Vancouver, British Columbia, V6T 1Z4 Canada}

\author{Matteo Michiardi}
\affiliation{Quantum Matter Institute, University of British Columbia, Vancouver, British Columbia, V6T 1Z4 Canada}

\author{Darius Menezes}
\affiliation{Department of Physics and Astronomy, University of British Columbia, Vancouver, British Columbia, V6T 1Z1 Canada}
\affiliation{Quantum Matter Institute, University of British Columbia, Vancouver, British Columbia, V6T 1Z4 Canada}

\author{Jiabin Yu}
\affiliation{Department of Physics and Astronomy, University of British Columbia, Vancouver, British Columbia, V6T 1Z1 Canada}
\affiliation{Quantum Matter Institute, University of British Columbia, Vancouver, British Columbia, V6T 1Z4 Canada}

\author{Sergey Zhdanovich}
\affiliation{Quantum Matter Institute, University of British Columbia, Vancouver, British Columbia, V6T 1Z4 Canada}

\author{Andrea Damascelli}
\affiliation{Department of Physics and Astronomy, University of British Columbia, Vancouver, British Columbia, V6T 1Z1 Canada}
\affiliation{Quantum Matter Institute, University of British Columbia, Vancouver, British Columbia, V6T 1Z4 Canada}

\author{David J. Jones}
\affiliation{Department of Physics and Astronomy, University of British Columbia, Vancouver, British Columbia, V6T 1Z1 Canada}
\affiliation{Quantum Matter Institute, University of British Columbia, Vancouver, British Columbia, V6T 1Z4 Canada}

\author{Sarah A. Burke}
\affiliation{Department of Physics and Astronomy, University of British Columbia, Vancouver, British Columbia, V6T 1Z1 Canada}
\affiliation{Quantum Matter Institute, University of British Columbia, Vancouver, British Columbia, V6T 1Z4 Canada}
\affiliation{Department of Chemistry, University of British Columbia, Vancouver, British Columbia, V6T 1Z1 Canada}

\date{\today}

\begin{abstract}
As an organic semiconductor and a prototypical acceptor molecule in organic photovoltaics, C$_{60}$ has broad relevance to the world of organic thin film electronics. Although highly uniform C$_{60}$ thin films are necessary to conduct spectroscopic analysis of the electronic structure of these C$_{60}$-based materials, reported C$_{60}$ films show a relatively low degree of order beyond a monolayer. 
Here, we develop a generalizable two-stage growth technique that consistently  produces single-domain C$_{60}$ films of controllable thicknesses, using Au(111) as an epitaxially well-matched substrate. We characterize the films using low-energy electron diffraction, low-energy electron microscopy, scanning tunneling microscopy, and angle-resolved photoemission spectroscopy (ARPES). We report highly oriented epitaxial film growth of C$_{60}$/Au(111) from 1 monolayer (ML) up to 20~ML films. The high-quality of the C$_{60}$ thin films enables the direct observation of the electronic dispersion of the HOMO and HOMO-1 bands via ARPES without need for small spot sizes. Our results indicate a path for the growth of organic films on metallic substrates with long-range ordering.
\end{abstract}
\maketitle  
\noindent  

\section{Introduction}

Organic semiconductors (OSCs) offer intriguing alternatives to conventional semiconductor devices owing to the possibility of tuning their properties through chemical modification, their flexible and light-weight nature, and their low production cost \cite{Kaltenbrunner:2012, Xu:2018, Gambhir:2016}. These advantages have led to the development of organic thin film transistors (OTFTs), organic light-emitting diodes (OLEDs), and organic photovoltaic devices (OPVs) \cite{Dimitrakopoulos:2002, Swayamprabha:2021, Brabec:2001}. For OPVs, lower energy production and unique form-factors hold particular economic and societal appeal \cite{Service:2022}. However, issues with disorder in organic semiconductors make the intrinsic electronic structure challenging to characterize and engineer. Molecular films often exhibit a high degree of disorder when compared to their inorganic counterparts due to the soft potentials, multiple crystal structures, and weak epitaxy with typical substrates \cite{Hooks2001}. These inhomogeneities in the sample confounds measurements of electronic states \cite{Ginsberg:2015, Silva:2011}. Here, we use C$_{60}$ as a test case for the growth of highly ordered organic films due to its relative structural simplicity, well-matched epitaxy with Au(111), and featured role as a strong acceptor in OPVs.

 In order to probe the electronic band structure of C$_{60}$-based materials, high purity films with long-range order and atomically-consistent thicknesses exceeding a monolayer (ML) are required \cite{Latzke:2019, Stadtmuller:2020}. Molecular beam epitaxy (MBE) is the technique of choice for achieving high quality films of inorganic materials with atomic-layer control and is similarly used to grow C$_{60}$ films and related small-molecule semiconductors. Despite the use of MBE and MBE-like growth approaches, techniques requiring a high degree of orientational order like angle-resolved photoemission spectroscopy (ARPES) have been impeded by the effects of such mosaicity, requiring measurements from very small spot sizes like those at synchrotrons or with momentum microscopes, reducing their accessibility. Additionally, while not all measurement techniques are directly impacted by disorder, the role of interfaces in disrupting charge transport can impact device performance \cite{Ginsberg:2015}.

\begin{figure}[ht!]
	\centering
	\includegraphics{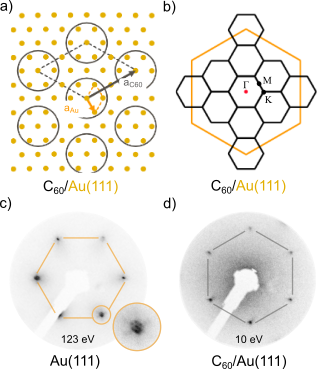}
	\caption{a) Schematic of one of C$_{60}$'s preferential epitaxial growths on Au(111) -- a $2\sqrt{3}$ x $2\sqrt{3}$ R30$^{\circ}$ superstructure. Unit cells are delineated with dashed lines, and lattice vectors are solid arrows. b) Schematic of epitaxial relationship between C$_{60}$ BZ in black and Au(111) BZ in orange. c) LEED image of Au(111) substrate prepared for growth. Inset shows LEED image of herringbone structure of clean Au(111).\cite{Haag:2016} d) LEED image of ML C$_{60}$ on Au(111), grown with stage 1 process. Note the 30$^{\circ}$ rotation in the orientation of the C$_{60}$ diffraction peaks relative to that of the gold substrate in panel c.}
	\label{fig:fig1}
\end{figure}

Here, we describe a two-stage growth recipe, combining hetero-epitaxial and homo-epitaxial growth processes at different temperatures. 

\begin{itemize}
    \item Stage 1 (Heteroepitaxial Growth): Deposit $>$ 1~nm of C$_{60}$ at a rate between 0.01 and 0.1~nm/min, maintaining a substrate temperature of 300$^{\circ}$ C, above the re-evaporation temperature for additional layers, promoting high mobility, and preventing growth beyond 1~ML. 
    \item Stage 2 (Homoepitaxial Growth): Deposit C$_{60}$ at a rate between 0.01 and 0.1~nm/min, maintaining a substrate temperature of 155$^{\circ}$ C (below the temperature for re-evaporation but still promoting mobility), until desired film thickness has been achieved. 
\end{itemize}

This process enables consistent growth of high purity C$_{60}$ films on Au(111) in the $2\sqrt{3}$ x $2\sqrt{3}$ R30$^{\circ}$ superstructure by holding the substrate at an elevated temperature. A schematic of the $2\sqrt{3}$ x $2\sqrt{3}$ R30$^{\circ}$ structure is shown in Fig.\,\ref{fig:fig1}a and the epitaxial relationship between the C$_{60}$ Brillouin zone (BZ) and Au(111) BZ are shown in Fig.\,\ref{fig:fig1}b. We confirm orientation of the growth using low energy electron diffraction (LEED). Figure \ref{fig:fig1}c and d show LEED images of the clean Au(111) substrate and ML C$_{60}$, respectively. The diffraction patterns are rotated 30$^{\circ}$ with respect to each other, confirming the growth is in the $2\sqrt{3}$ x $2\sqrt{3}$ R30$^{\circ}$ structure.



\begin{figure}[t]
	\centering
	\includegraphics{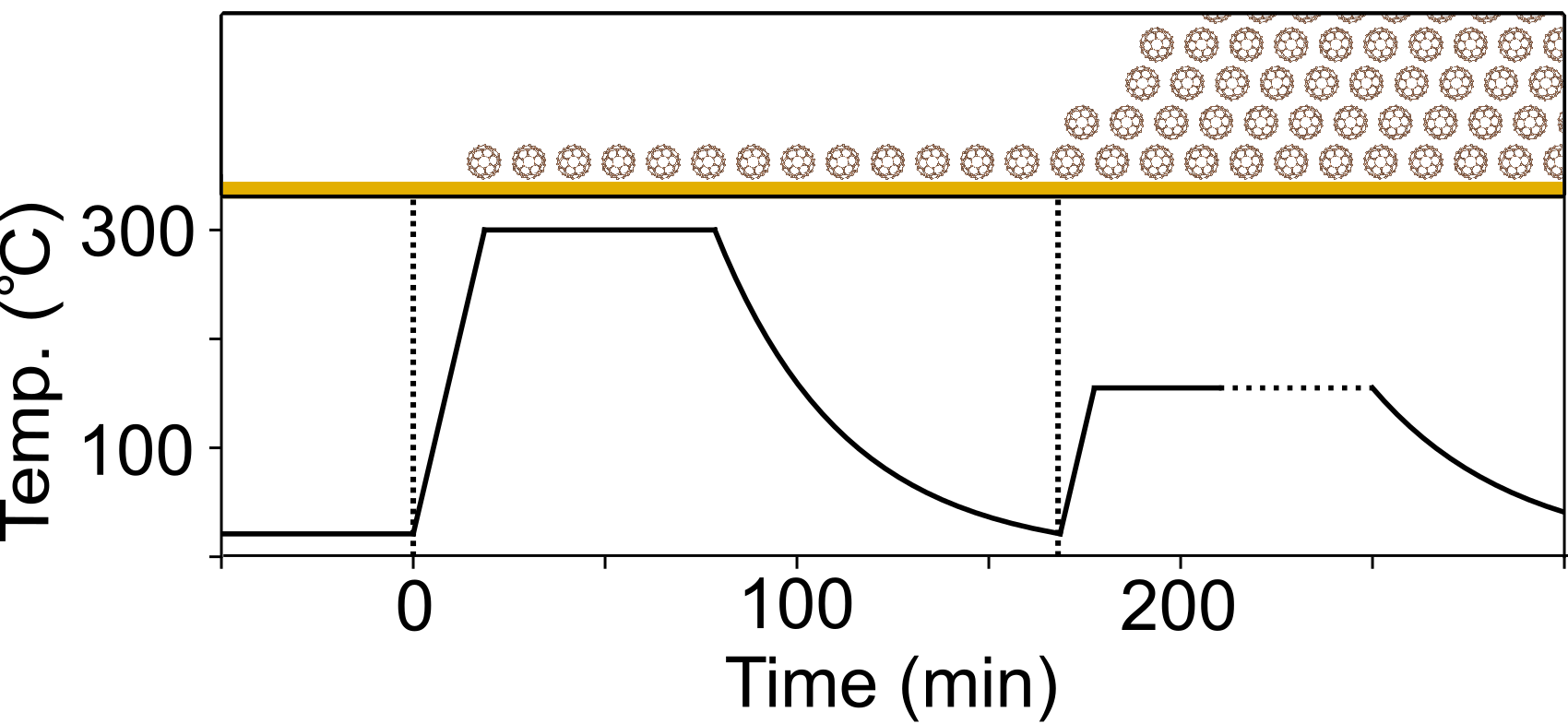}
	\caption{Illustration of two-stage film growth procedure, beginning with a clean substrate, following by growth of the first ML, and finally growth of the desired number of additional layers. Note that the first ML is grown at 300$^{\circ}$ C, and the subsequent layers are grown at 155$^{\circ}$ C.}
	\label{fig:C60growth}
\end{figure}

This two-stage growth relies on a high temperature over-deposition of 1~ML to produce a highly ordered monolayer film, similar to the distillation-like growth first shown to yield high quality ultraviolet photoemission spectroscopy \cite{Veenstra:2002, Sawatzky:1997} in which the monolayer C$_{60}$ was achieved by annealing the sample post-deposition at temperatures as low as 240$^{\circ}$ C.   The second stage follows as a homoepitaxial growth at  lower temperature templated by the ordered first layer (Fig.\,\ref{fig:C60growth}). The temperature of the substrate should be below 200$^{\circ}$ C for the homoepitaxial stage. This growth method is possible due to the strength of the bond between fullerene and metallic substrate in C$_{60}$/Au(111): the adsorption interaction should not be considered pure van der Waals but rather chemisorption, where charge transfer and LUMO-metal state mixing accounts for the additional interaction strength \cite{Ohno:1991}. We characterize the films resulting from this two-stage growth method using LEED, scanning tunneling microscopy (STM), and ARPES, demonstrating the long-range order needed for area-integrated spectroscopic measurements.

\section{LEED and LEEM}
\begin{figure}
	\centering
	\includegraphics{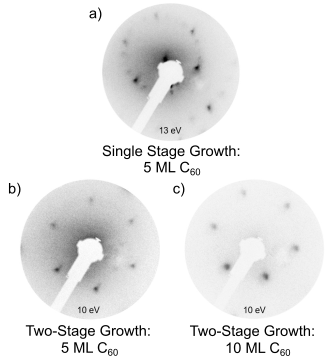}
    \caption{LEED data comparing the results of a typical one-stage growth recipe a) to films grown using our two-stage growth recipe b) and c). LEED images are taken with a beam energy of approximately 10~eV. a) 5~ML film grown with the substrate held at 155$^{\circ}$ C for the entire duration. LEED images show the presence of multiple domains. The LEED for the films grown using the two-stage method (see Fig.\ref{fig:C60growth}) with b) 5 ML and c) 10 ML shows no sign of multiple domains.}
	\label{fig:MD}
\end{figure}

Using LEED, we are able to identify the orientation of the overlayer unit cell and visualize the presence of multiple domains with a spot size of 250~$\upmu$m to  1~mm, this is a similar length scale to many lab based ARPES photon sources \cite{Moritz:2018}. Within the signal-to-noise limits of our LEED, our two-stage growth recipe repeatably generates uniform, well-ordered films up to at least 20~ML. As shown in Fig.\,\ref{fig:MD}, this recipe significantly improves previous efforts \cite{Haag:2020, Tzeng:2000, Altman:1992} to grow C$_{60}$ films thicker than 1~ML which used a single stage growth recipe with the substrate held at the second stage growth temperature ($\sim$ 155$^{\circ}$ C).
In LEED images, multiple rotational domains in the sample manifest as multiple single-domain LEED patterns rotated relative to one another. In the case of a C$_{60}$ film, the hexagonal pattern is rotated forming a ring of spots (Fig.\,\ref{fig:MD}a). The relative intensity of the spots gives an indication of the prominence of a  domain \cite{VanHove:1986}. Figure \ref{fig:MD}a depicts the LEED pattern of a 5~ML C$_{60}$ film grown using a single-stage growth with the temperature of the substrate held at 155$^{\circ}$ C. The $2\sqrt{3}$ x $2\sqrt{3}$ R30$^{\circ}$ domain is dominant with three other domains visible. Figure \ref{fig:MD}b is of similar thickness to panel a, but was grown using the two-stage method described in Fig.\,\ref{fig:C60growth}. This LEED image shows signs of only a single domain. Figure \ref{fig:MD}c is a 10~ML film grown using the two-stage method again showing signs of only a single domain. We also observe similar results in thicker films (at least 20~ML).

\begin{figure}
	\centering
	\includegraphics{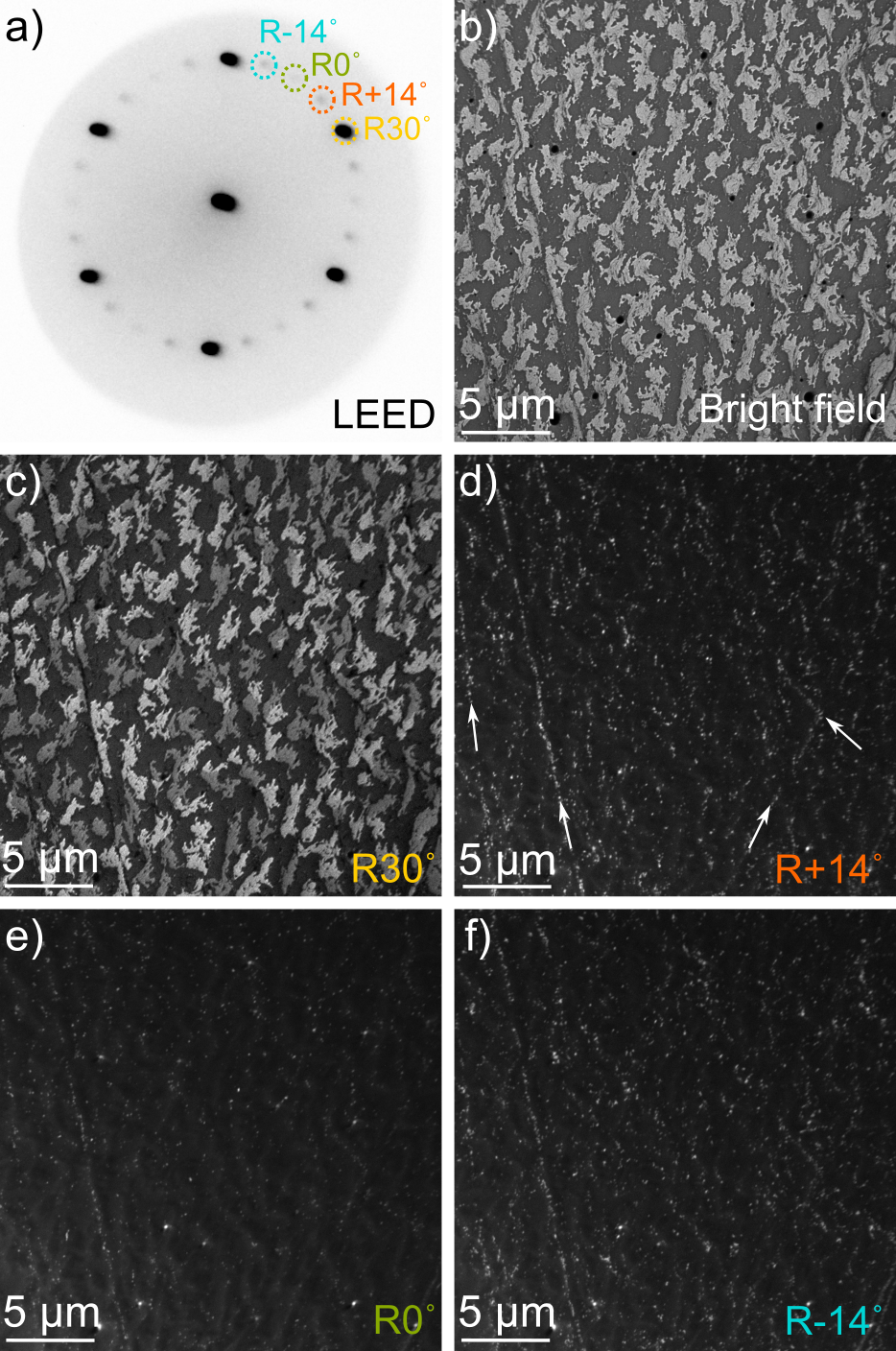}
    \caption{LEEM results of 1.5 ML C$_{60}$ single-stage growth showing multidomain diffraction. a) LEED image (Start Voltage, SV, 5 V) typical of single-stage growth recipes. b) bright-field LEEM image (SV=3.3 V) showing contrast between first layer and second layer. c-f) Dark-field LEEM images (SV=3.3 V) corresponding to R30$^{\circ}$ c), +14$^{\circ}$ d), in-phase e), -14$^{\circ}$ f) domains relative to the Au(111) lattice. Examples of locations where grains appear to be nucleated at step edges are indicated by arrows in d), including a 60$^{\circ}$ bend at the right-most arrow indicative of the underlying Au(111) surface. The two different shades of gray seen among the 2 ML islands in c) are due to a slight contrast difference between the +30$^{\circ}$ and -30$^{\circ}$ domains, which we attribute to interactions with the underlying Au(111) substrate. All images are 25~$\upmu$m x 25~$\upmu$m.}
	\label{fig:LEEM}
\end{figure}

To further investigate the origins of multi-domain growth, we use low-energy electron microscopy (LEEM) to visualize a 1.5~ML film grown in a single-stage at an intermediate temperature of 190$^{\circ}$ C. Figure \ref{fig:LEEM} shows the diffraction, bright-field and selected dark-field images for the resulting multi-domain film. While the $2\sqrt{3}$ x $2\sqrt{3}$ R30$^{\circ}$ domains (Fig.\,\ref{fig:LEEM}c) form the predominant structure at this growth temperature, there is notable diffraction intensity from two other domains corresponding to the in-phase (R0$^{\circ}$, Fig.\,\ref{fig:LEEM}e) and $7$ x $7$ R14$^{\circ}$ (R$\pm$14$^{\circ}$, Fig.\,\ref{fig:LEEM} d,f)) structures previously reported \cite{Gardener:2009}. The dark-field images corresponding to these domains (Fig.\,\ref{fig:LEEM}d-f) show small nucleation centres that appear to mostly be aligned along step edges (see arrows in Fig.\,\ref{fig:LEEM}d). This is consistent with the observations of Altman et al. \cite{Altman:1992} that found nucleation and growth of the in-phase structure from the step edges is kinetically favoured, despite the thermodynamically favoured $2\sqrt{3}$ x $2\sqrt{3}$ R30$^{\circ}$. 

\section{STM}
\begin{figure}
	\centering
	\includegraphics{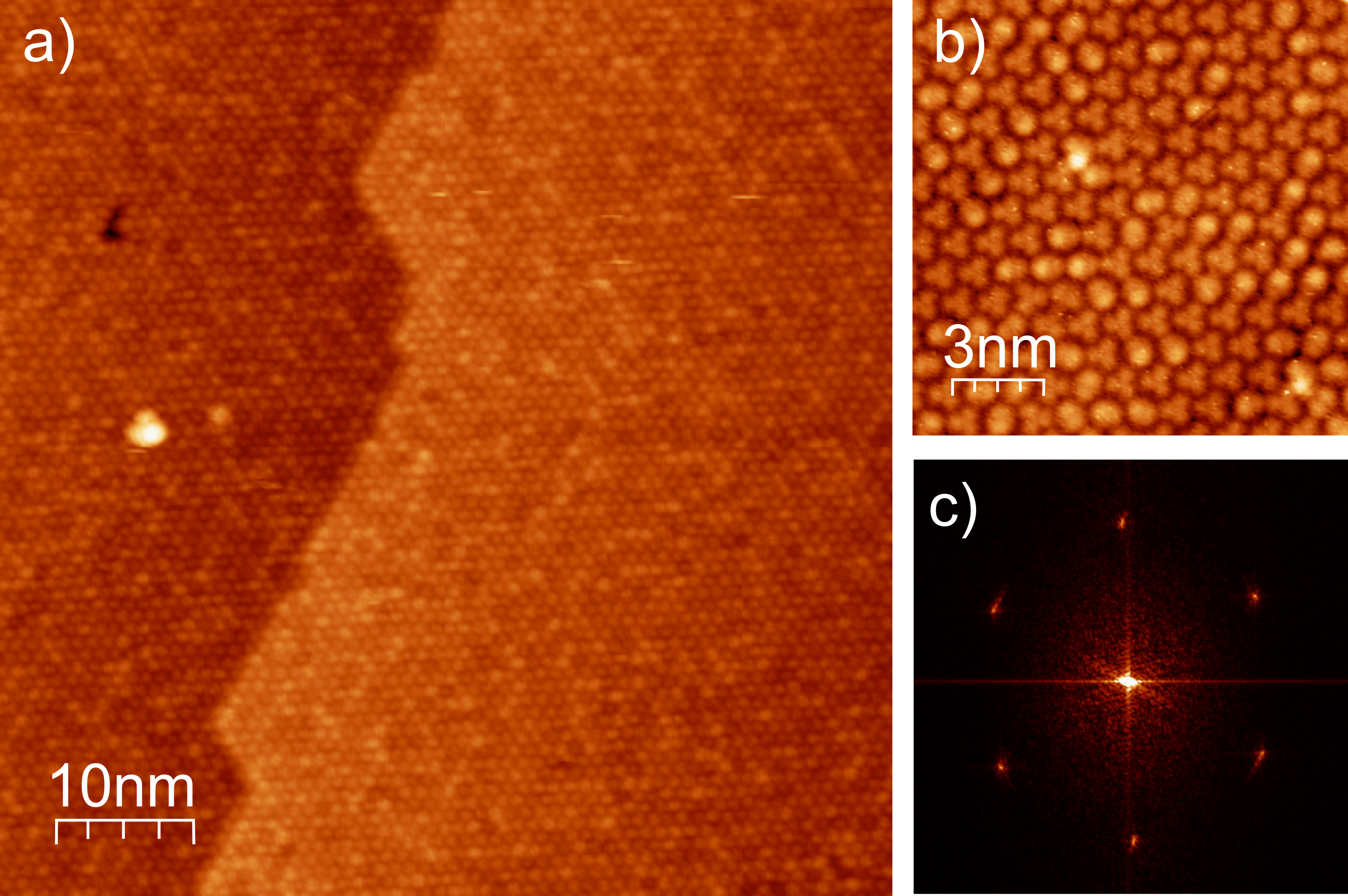}
	\caption{STM of 1 monolayer C$_{60}$/Au(111) grown using the first-stage of the growth recipe taken at 4.6~K. a) STM image showing the long range order of the C$_{60}$ overlayer ($V_\text{Bias}$ = -2.5~V, $I_\text{setpoint}$ = 5~pA, $\Delta z$ = 1.135~nm). b) STM image showing fine electronic structure of the C$_{60}$ ML taken on a different sample than a) ($V_\text{Bias}$ = -2.5 V, $I_\text{setpoint}$ = 5~pA, $\Delta z$ = 0.2596~nm). c) Fourier tansform of a) shows only single rotational domain.}
	\label{fig:STM}
\end{figure}

To study the quality and degree of order of the first-stage monolayer growth, we also employ low-temperature STM. Figure \ref{fig:STM}a depicts a typical STM image of a sample grown using the first-stage of the growth recipe and then transferred into the STM via a passive transfer chamber. The transfer takes $\sim$1 hour during which the sample is exposed to pressures between $1 \times 10^{-8}$ mbar and $4\times 10^{-5}$ mbar. Images of the C$_{60}$ film are consistent with prior STM results, showing bright and dim spots pertaining to C-C bond down and hexagon down on the Au(111) substrate (Fig.\,\ref{fig:STM}b) \cite{Gardener:2009, Shin:2014}. The film shows some evidence of minimal surface contamination from the transfer (e.g. one bright spot in Fig.\,\ref{fig:STM}a - typical for scan this size). The step edge visible through the center of the image and dark spot correspond to a step edge and a pit in the underlying Au(111) substrate, respectively. While step edges in the Au(111) substrate are observed, no steps consistent with a 2nd layer of C$_{60}$ are seen, indicating that the closed first ML is self-limiting and additional C$_{60}$ deposition leads to re-evaporation without further growth at this temperature.

A Fourier transform of the STM image (Fig.\,\ref{fig:STM}c) confirms a single rotational domain. To ascertain the uniformity and degree of order of the film, the Fourier transform of every STM image acquired for two different samples are compared, confirming that the C$_{60}$ overlayer is largely unaffected by minor defects in the substrate. We find that $>97\%$ of randomly selected scanning areas (total area 0.48~$\upmu$m$^2$ across the 8~mm diameter crystal) exhibit the same rotational domain. The few areas showing multiple domains are found in regions of the sample with a large number of steps in the substrate (eg. Fig.S\,1), that are also typically near the edge of the crystal. This is consistent with the LEEM data and supports the need for good substrate preparation in addition to the two-stage growth procedure.


Undergrown samples are also examined by STM. These samples are grown using the same parameters as the first-stage of the growth recipe to a coverage of $80\%$ of a ML rather than over-depositing. In these cases ordered C$_{60}$ is observed in $\sim78\%$ of the randomly selected scanning area as anticipated, while the other $\sim22\%$ is disordered (eg. Fig.S\,2), possibly consisting of low-density C$_{60}$ and Au adatoms from the lifted reconstruction and expected to be mobile at this growth temperature. This highlights the importance of over depositing a ML in the first-stage of the growth recipe to reduce disorder in this foundation layer. The STM measurements are summarized in table 1 of the supplementary information. 



\section{ARPES}


\begin{figure}
	\centering
	\includegraphics{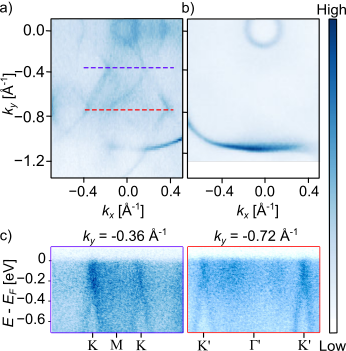}
  	\caption{ARPES measurements of the substrate and first-stage growth. a) Fermi surface (FS) after the first-stage growth; Replicas of the Au s-p band are created by the C$_{60}$ overlayer. These measurements were performed at 10~K using a HHG laser at 23~eV and 100~$\upmu$m spot size. b) FS of the clean Au(111) substrate measured at 10~K using a helium lamp with 21.2~eV photons. c) Dispersion of the reconstructed bands at the high symmetry cuts indicated in panel a).}
  
	\label{fig:AuML}
\end{figure}

With long-range order confirmed by both LEED and STM, we now demonstrate characterization of the electronic structure of these C$_{60}$/Au(111) films by ARPES. The Fermi surface (FS) of the ML C$_{60}$ films, formed by the stage 1 heteroepitaxial growth, is shown in Fig.\,\ref{fig:AuML}a, with the FS of the clean Au(111) substrate over the same momentum range shown in Fig.\,\ref{fig:AuML}b for reference. The Au(111) FS shows both the surface-resonance s-p band at high momenta and the Shockley surface state at the center of the BZ, both of which cross the Fermi level. In the ML C$_{60}$ FS, the Au(111) surface-resonance s-p band is still visible, in addition to new features following the expected C$_{60}$ BZ. However, as C$_{60}$ is a semiconductor, it is not expected to contribute any bands that cross the Fermi level. Instead, these additional bands correspond to reconstructions of the surface-resonance Au(111) s-p band formed by the new periodicity imposed by the C$_{60}$ overlayer. These replica bands form quasi-linear crossings reminiscent of Dirac cones at the K points, as shown in Fig.\,\ref{fig:AuML}c. These spectral features have been observed previously, however, the origin of the reconstruction -- whether it is bandfolding or photoelectron diffraction
-- is still a subject of debate \cite{Krivenkov:2022, Yue:2020}. Regardless of its origin, the presence of the reconstruction further confirm the high degree of orientational order in the first C$_{60}$ monolayer. Features associated with the Au(111) dominate the spectra, while the HOMO and HOMO-1 of C$_{60}$ are less prominent and appear at lower binding energy compared to thicker films (see Fig.S\,4). The ML C$_{60}$ measurements included in Fig.\,\ref{fig:AuML}a and c were performed using a HHG laser at 23~eV and 100~$\upmu$m spot size \cite{Mills:2019}. However, these measurements were also performed using a He I line from a helium discharge lamp with 21.2~eV photons and a 1~mm spot size and the replica s-p bands are maintained (see Fig.S\,5), demonstrating that the films are suitable for ARPES measurements with a variety of lab-based photon sources.

\begin{figure*}
    \includegraphics[width=1.0\textwidth]{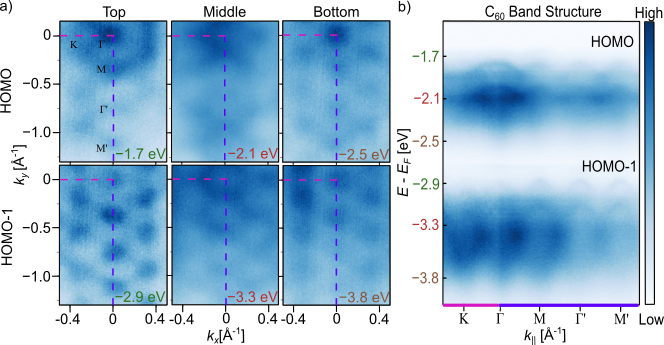}
    \caption{ARPES measurements of 5~ML C$_{60}$/Au(111) grown using the two-stage growth recipe. a) Constant energy contours (CEC) of the HOMO (top row) and HOMO-1 (bottom row); no rotational domains (manifested as rotated replicas) are observed. b) Dispersion of the HOMO and HOMO-1 along the high-symmetry directions outlined in panel a). Measurements were performed at 10~K using a table-top HHG laser with photon energy 23~eV and spot size 100~$\upmu$m.}
	\label{fig:5ML}
\end{figure*}

We now describe the ARPES results on the two-stage multilayer film, shown in Fig.\,\ref{fig:5ML}. The film thickness for this dataset is 5~ML, as determined by quartz crystal monitors during the second-stage deposition. The C$_{60}$ HOMO and HOMO-1 are clearly resolved at binding energies of 2.1~eV and 3.3~eV, with a bandwidth of 550 meV and 780 meV, respectively. Constant energy contours (CEC) at the top, middle, and bottom of the HOMO and HOMO-1 are shown in the top and bottom row of Fig.\, \ref{fig:5ML}a, respectively, clearly showing the presence of one dominant orientation with no evidence of other rotational domains. CECs from data acquired with a He I line from a helium discharge lamp, with 21.2~eV photons, of a 20~ML film show similar features and data quality (see Fig.S\,6) indicating thicker films maintain the long range order seen in the first ML. The dispersion along the high symmetry directions indicated by the purple and magenta dashed lines is shown in Fig.\,\ref{fig:5ML}b. The ARPES measurements were done at 10~K, below the orientational phase transitions of C$_{60}$ corresponding to the simple cubic phase \cite{Heiny1992}. Our results are consistent with previous reports using small spot sizes and with prior calculations of the C$_{60}$ band structure in the simple cubic phase\cite{Laouini:1995, Latzke:2019, Haag:2020}.

 
\section{Conclusion}
Here we have demonstrated a two-stage recipe that exploits both homoepitaxial and heteroepitaxial growth, resulting in highly ordered films confirmed by LEED and STM compared to those grown using a more traditional, single-stage recipe. Such films enable techniques such as ARPES, that require long range orientational order, allowing us to make accurate measurements of C$_{60}$'s electronic structure. This approach exploits a strong molecule-substrate interaction and may be generalizable to other organic semiconductors which often interact strongly with metals through $\pi$-bonding. The preparation of such long-range ordered films enables the use of ARPES with lab-based, large-spot photon sources to investigate the electronic structures of organic thin films, taking a core technique for the study of electronic structure determination in single crystals over to the realm organic films. The high uniformity of films demonstrated here - lacking in grain boundaries, unintended interfaces - and other defects, is also beneficial for other spectroscopic techniques and devices, with potentially broad implications for OLEDs, OTFTs, and other organic semiconductor-based technologies.


\section{Methods} 
The C$_{60}$ films are grown in an MBE chamber with a base pressure of $7.4\times10^{-11}$ Torr. The C$_{60}$ is evaporated using a water-cooled quad cluster source from MBE Komponenten with PBN crucibles, and the growth rate is calibrated using a water-cooled quartz crystal monitor. Constant growth rate is maintained throughout each of the deposition; deposition rates ranging from 0.01 to 0.1~nm/min were used. The temperature of the first-stage homoepitaxial growth was held at 300 $^{\circ}$ C and a first ML was over deposited. The second-stage heteroepitaxial growth was held at 155 $^{\circ}$ C. In order to ensure consistent $2\sqrt{3}\times2\sqrt{3}R30^{\circ}$ superstructures, excellent substrate preparation is necessary. We consider the Au(111) substrate well prepared when the herringbone reconstruction is visible in the LEED pattern.  

The Au(111) single-crystal substrate is prepared by sputtering with Ar+ with a SPECS’ IQE 11/35 ion source held at 1.3 keV for 40 minutes to 1 hour, followed by annealing the substrate at 440$^{\circ}$ C for 90 minutes. We repeat this sputtering and annealing process if necessary, to ensure the substrate is clean and flat. We confirm the substrate is properly prepared using LEED. Although LEED images of the Au(111) indicate the presence of a herringbone reconstruction (see Fig.\,\ref{fig:fig1}c), it is well-documented in the literature that C$_{60}$ thin-films lift the herringbone reconstruction in gold \cite{Altman:1992}.
Once the substrate is prepared, we begin our two-stage growth process. The C$_{60}$-Au bond can withstand 500$^{\circ}$ C temperatures without C$_{60}$ desorption, whereas the C$_{60}$-C$_{60}$ van der Waals bonds break down in temperatures above 300$^{\circ}$ C \cite{Altman:1993}. Thus, in stage 1, when our substrate is held at 300$^{\circ}$ C, despite depositing nearly 3~nm of C$_{60}$ onto the substrate, C$_{60}$ layers beyond the first ML are desorbed and we are left with a clean, well-ordered ML as determined by STM and ARPES. 


The LEEM measurements were carried out at the MAXPEEM beamline of the MAX IV synchrotron in Lund, Sweden. A Au(111) crystal was prepared through sputtering and annealing cycles in a separate preparation chamber. C$_{60}$ was deposited from a home-made dual crucible evaporator directly in the LEEM chamber during live LEEM imaging. The evaporator was mounted at a 16$^{\circ}$ angle to the sample surface and the deposition rate was approximately 0.01~nm/min. The base pressure in the LEEM chamber during deposition was $3\times10^{-9}$ mbar.

All STM measurements were taken in a ScientaOmicron low-temperature scanning probe microscope with a base chamber pressure of $6\times10^{-11}$ Torr, using a cut platinum-iridium tip. All measurements were taken at 4.6~K. Samples were transferred from the MBE chamber to the STM UHV system in a passive vacuum chamber, and were exposed to pressures between $1 \times 10^{-8}$~mbar to $4 \times 10^{-5}$~mbar. Transfers took $\sim$1 hour to complete. 

The MBE chamber is attached directly to both a LEED chamber and an ARPES system, facilitating in situ sample growth and analysis. The LEED is a BDL600IR-MCP model from OCI Vacuum Microengineering Inc., with a 7 - 750~eV beam energy and a spot size of approximately 250~$\upmu$m. The base chamber pressure is $1.5\times10^{-10}$~Torr. The ARPES system contains a Scienta DA30L hemispherical electron analyzer with an XUV light source. The XUV source has been modified from that previously reported \cite{Mills:2019} to provide a tunable photon energy (8.5-35 eV) and smaller spot size (approximately 100 um).  The output is $\pi$-polarized, and at a photon energy of 23 eV has an energy resolution of 40 meV. 
The base chamber pressure is $5\times10^{-11}$~Torr.




The authors thank discussion and support from George A. Sawatzky, Jisun Kim, James Day and  Hsiang-Hsi Kung.

This research was undertaken thanks in part to funding from the Max Planck-UBC-UTokyo Centre for Quantum Materials, and the Canada First Excellence Research Fund in Quantum Materials and Future Technologies Program. This project is also funded by the Gordon and Betty Moore Foundation's EPiQS Initiative, Grant No. GBMF4779 to A.D. and D.J.J.; the Natural Sciences and Engineering Research Council of Canada (NSERC); the Canada Foundation for Innovation (CFI); the British Columbia Knowledge Development Fund (BCKDF); the Department of National Defense (DND); the Canada Research Chairs Program (S.A.B. and A.D.); the CIFAR Quantum Materials Program (A.D.); and the NSERC PGSD and Tyler Lewis Clean Energy Research grants program (A.T.). We acknowledge MAX IV Laboratory for time on Beamline MAXPEEM under Proposal 20210366. Research conducted at MAX IV, a Swedish national user facility, is supported by the Swedish Research council under contract 2018-07152, the Swedish Governmental Agency for Innovation Systems under contract 2018-04969, and Formas under contract 2019-02496.


\bibliography{bibliography}

\end{document}